\documentclass[english]{lni}
\let\ifpdf\relax

\usepackage[utf8]{inputenc}
\usepackage{url}
\usepackage{graphicx}
\usepackage{todonotes}
\usepackage{authblk}

\newcommand\protbox{Protbox}
\newcommand\ProtF{Prot Folder}
\newcommand\ProtFs{Prot Folders}
\newcommand\SharedF{Shared Folder}
\newcommand\SharedFs{Shared Folders}
\newcommand\CloudF{Cloud Folder}
\newcommand\CloudFs{Cloud Folders}
\newcommand\PairKey{Pair Key}
\newcommand\PairKeys{Pair Keys}
\newcommand\PPair{{\protbox} Pair}
\newcommand\PPairs{{\protbox} Pairs}

\begin{document}

\author[1]{
	Eduardo Duarte,
    emod@ua.pt
}
\author[2]{
	Filipe Pinheiro,
    filipepinheiro@ua.pt
}
\author[3]{
	André Zúquete,
    andre.zuquete@ua.pt
}
\author[4]{
	Hélder Gomes,
    helder.gomes@ua.pt
}
\affil[1]{University of Aveiro}
\affil[2]{University of Aveiro}
\affil[3]{DETI/IEETA, University of Aveiro}
\affil[4]{ESTGA/IEETA, University of Aveiro}

\title{Secure and trustworthy file sharing over cloud storage using eID tokens}

\maketitle
\begin{abstract}
This paper presents a multi-platform, open-source application that aims to
protect data stored and shared in
existing cloud storage services. The access to the cryptographic
material used to protect data is implemented using the
identification and authentication functionalities of
national electronic identity (eID) tokens.
All peer to peer dialogs to exchange cryptographic material is implemented
using the cloud storage facilities. 
Furthermore, we have included a set of mechanisms to prevent files
from being permanently lost or damaged due to concurrent modification,
deletion and malicious tampering.

We have implemented a prototype in Java that is agnostic relatively
to cloud storage providers; it only manages local folders, one of
them being the local image of a cloud folder. We have successfully
tested our prototype in Windows, Mac OS X and Linux, with Dropbox, 
OneDrive, Google Drive and SugarSync.
\end{abstract}

\section{Introduction}

In recent years we assisted a widespread usage of cloud storage for
centrally storing personal files (e.g. Dropbox). Such cloud storage
can either be used for personal benefit or for sharing information
with others. In this last case, cloud storage providers manage the
mechanisms to send sharing invitations and to keep the shared files
synchronized among all the hosts effectively using them. To ease the
usage of such shared folders, storage providers enable users to use
in their hosts specific software to handle a mount point in the
local file system to access cloud folders.

In this paper we propose a system, \textbf{\protbox}, for securely
sharing files among strongly authenticated people through many different cloud
storage services. The secure sharing includes four different
protection features: (i) confidentiality, to prevent
non-authorized readings, (ii) integrity control, to detect malicious
tampering, (iii) protection against unwanted file removals,
either by malicious or legitimate persons, and (iv)
access control to the shared
data based on strong identification and authentication
of people, using the nowadays widespread electronic,
personal identity tokens (eIDs for short).

Many governments worldwide have been or are introducing eIDs to allow
the identification of people in the scope of Internet interactions.
Unfortunately, there are several kinds of eID types being deployed,
which reduces the possibilities of using all of them in a single
system requiring the authentication of persons. In our system we
considered the case of the Portuguese eID (Cartão de Cidadão), which
enables the owner to perform two kinds of signatures upon providing
a proper PIN: (i) authentication signature, for online identity
proofs and (ii) qualified signatures, for document signing. In this
work we used only authentication signatures.

Comparing {\protbox} with similar solutions, it has two main distinctive
characteristics: (i) the key distribution between file sharing
persons is performed by means of special files exchanged through the
exact same cloud storage space used for file sharing, thus no extra services
are required other than the trustworthy national PKIs (Public Key
Infrastructures) used to validate eID signatures; and (ii) the files
exposed to others by means of cloud sharing are protected from
malicious or involuntary tampering or removal.

{\protbox} has just two requirements regarding a cloud storage
solution for folders and files: (i) it should allow the sharing of  folders by many persons and (ii) it should allow client
operating systems to have a local mount point of the shared folder.
Nowadays, most file-oriented cloud storage solutions, if not all,
fulfill these requirements; in our experiments we managed to explore
it successfully with Dropbox, SkyDrive, Google Drive and SugarSync
(see Section~\ref{experiences}).

We developed a {\protbox} prototype in Java. It runs in any
operating system with a suitable Java Virtual Machine (JVM) and is
capable of recognizing any file system. It features a background
folder synchronization engine and a graphical user interface for
dealing with key distribution requests. {\protbox} randomly
generates and uses a key per folder to protect all its contents,
including files and sub-directories. Files are encrypted with AES and their integrity is ensured with
HMAC-SHA1. Encrypted file names, which contain bytes that are not
acceptable for naming files in existing file systems, are coded in a
modified Base64 alphabet, which should work in most file systems.
The prototype was successfully experimented in Windows, Mac OS X and 
Linux with all of the above referred cloud storage providers.





\section{{\protbox} architecture}

\subsection{Deployment overview}

{\protbox} depends on the local replica of \textbf{{\CloudF}},
which we call \textbf{\SharedFs} for the effective sharing of
protected content. Users must define one-to-one associations between
those {\SharedFs} and the local folders containing the relevant
files to protect, which we call \textbf{\ProtF} and may be
located anywhere in the host file system. The cloud storage system
will be responsible for synchronizing the contents of {\SharedFs}
with the correspondent {\CloudFs}, which may be shared by a set
of cloud storage users. This cloud synchronization is completely
transparent to {\protbox}, which only has to deal with the local
synchronization between associated {\SharedF} - {\ProtF} pairs
(see Figure~\ref{{\protbox}.Fig}). We will refer such pairs as
\textbf{{\PPairs}}.

\begin{figure}[t]
\includegraphics[width=\columnwidth]{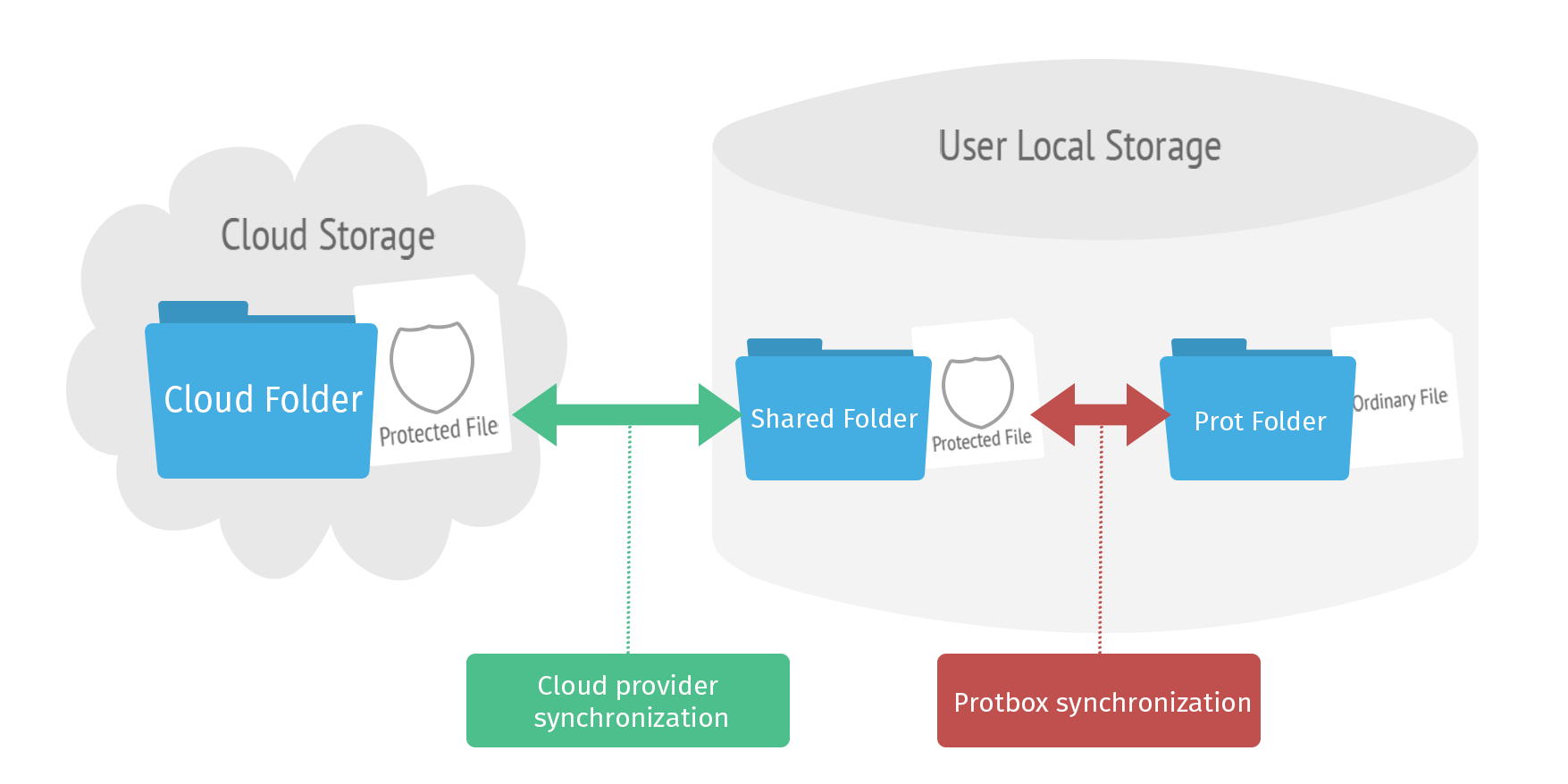}
\caption{Overview of the {\protbox} deployment architecture. The
{\protbox} synchronization will take place upon {\PPairs}, which are pairs formed by one {\SharedF} and a {\ProtF}.}
\label{{\protbox}.Fig}
\end{figure}

The {\protbox} synchronization actions are not simple file copies,
but rather content encryptions (when updating files in a {\SharedF})
or decryptions and integrity validations (when updating a
{\ProtF}). Files in a {\ProtF} are in their original formats,
as produced by the creating applications, but their protected
replicas in {\SharedFs} and {\CloudFs} are encrypted, ensuring
their confidentiality. The file names of the encrypted files are
also encrypted to reduce the leakage of file-related information
through the cloud provider.

\subsection{Integrated file protection}

{\protbox} attempts to introduce and build a confidential,
trustworthy and dependable environment on top of existing cloud
storage services without disrupting normal usage and functionality
provided by these.

For confidentiality, {\protbox} encrypts files with a symmetric
cipher (e.g. AES).  A key is generated and maintained for each
{\PPair} (hereafter we will call it a \textbf{{\PairKey}}).
{\PairKeys} are stored by {\protbox} in its private, local data repositories
and not stored in the cloud. Each {\PairKey} is either (i) randomly
generated by {\protbox} (when the first protected file is created in
the corresponding {\SharedF}) or (ii) imported by {\protbox} from
other users sharing the corresponding {\SharedF} (when the {\SharedF}
is not empty). Besides file's contents, their names are also
encrypted with the same key and written in a modified Base64 alphabet.

For trustworthiness, file updates in {\SharedFs} must also be
validated taking the corresponding {\PairKeys} into consideration: an
update can only be accepted if it was made by someone knowing the
correct {\PairKey}.  Otherwise, tampered files in {\SharedFs}
could originate files with garbage contents in {\ProtFs}.
Therefore, protected files contain a cryptographic checksum,
computed with their {\PairKey} (e.g. with HMAC-SHA1).

Considering a hierarchical rank of power, a {\ProtF} outranks its
corresponding {\SharedF}, thus a {\ProtF} file cannot ever be
permanently damaged or deleted upon a synchronization event
originated from the cloud provider (regardless of the ultimate
origin of the event). Therefore, for dependability the files in
each {\ProtF} should always be available for restoring at any
time, regardless of the cloud provider's interpretation of the
files' status.

The coherence of the files in both of these folders ({\ProtFs}
and {\SharedFs}) is assured by maintaining a parallel
control data structure containing structural information about both
folders (files and directories, encrypted and decrypted names, last
modified dates, lengths of contents). Coherency checking and
synchronization tasks would run on a periodic basis and use that
structural information and the effective contents of each {\protbox}
Pair to take the appropriate data transfer decisions. Hereafter we
will refer to this structural information as \textbf{{\protbox}
Registry} (\textbf{PReg} for short). {\PairKeys} are also part of PReg.

Note that a PReg is a local, private data structure that helps
a local {\protbox} instance to take the appropriate, local
decisions regarding file synchronizations, encryptions/decryptions
and recovery actions. In particular, a PReg is never synchronized
with another one.

Because we are essentially dealing with asynchronous copies of files
from one folder to another, with encryption and decryption of
contents and file names depending on which folder the file is
placed, concurrent file update conflicts can occur. These file
conflicts can be detected because the synchronization method uses the
PReg to evaluate each situation prior to the synchronization itself.
On
each run of the coherence checking task we create a index of the
files updated (i) only in the {\ProtF}, (ii) only in the {\SharedF}
and (iii) in both folders. The last ones are our subject of
interest in what concerns conflicts. Consequently, they will be synchronized in a different way, which
will lead to the production of two versions of the updated file,
while the other files will be synchronized in a straightforward way.
This feature also adds dependability to {\protbox}, because it
assures that conflicting updates are never destructive.

\subsection{Agnosticism and autonomy towards cloud storage providers}

Some cloud storage solutions do not provide any cryptographic
measures to protect the files they store (e.g. Microsoft's
OneDrive\footnote{\url{http://answers.microsoft.com/en-us/onedrive/wiki/sdfiles-sdperms/onedrive-and-data-encryption-is-your-data-secured/43ff303b-a6aa-4f02-8c47-b547d6a5ef14}}).
Other solutions implement security mechanisms to back up and encrypt
files, both in transit and at rest (e.g.
Dropbox\footnote{\url{https://www.dropbox.com/help/27/en}}), rendering the
service
HIPAA-compliant\footnote{\url{http://onr.com/secure-server-hosting/what-is-hipaa-compliance/}}.
However, they cannot guarantee that stored files are only decrypted
by user request, since the symmetric encryption keys used are
managed by the cloud storage providers.  To emphasize the safeguard
of files, some providers claim they have strict privacy policies
that prohibit company's employees from viewing the content of
stored files (e.g.
Dropbox\footnote{\url{https://www.dropbox.com/privacy}}), but while this
may be a deterrent measure, it does not effectively prevent it from
happening\footnote{\url{https://www.pcworld.com/businesscenter/article/260254/dropbox_gets_a_black_eye_in_spam_attack.html}}.

By realizing this, we designed {\protbox} as an agnostic solution
regarding cloud storage providers, being independent of both
(i) how they store and transfer files and (ii) how they implement
confidentiality and authentication mechanisms~\cite{cloudsecurity}. 
For {\protbox} the
only requirements from cloud storage providers are the availability
of what we called a {\SharedF}, a replica available in the local file
system of a {\CloudF}, and the synchronization of contents between
several {\SharedFs} for the same {\CloudF}.

Other than the cryptographic methods adopted in order to establish a
confidential environment, it was equally important that {\protbox}
allowed users to share protected files with each other. Access to the
original contents of protected files by a user should be controlled
within {\protbox}, regardless of who effectively has access to the
{\CloudF}, as determined by the cloud-storage provider. To do
so, {\protbox} implements request-response dialogs between
{\protbox} instances for exchanging {\PairKeys} associated to
{\SharedFs}. These dialogs are implemented with special files
stored in {\SharedFs}. When a {\protbox} instance first establishes an
association between a {\ProtF} and a {\SharedF}, if the {\SharedF}
is populated with files then it sends a request to
obtain the key to decrypt them. Such a request will be available to all
{\protbox} instances with {\ProtFs} associated with replicas of
that {\SharedF}, and any of them (upon a user consent) may send a
response with the requested key. This way, {\protbox} is completely
autonomous regarding key distribution, it does not require any external key
distribution service.

\subsection{Synchronization of {\PPairs}}

{\protbox} allows the user to configure a arbitrary number of local
{\ProtFs} to be securely shared by means of {\PPairs}.
This means that it must be able to properly synchronize data between
the {\ProtF} and the {\SharedF} that form each {\protbox}
Pair. In addition, while useful metadata provided by native file
systems can be used to detect updates and synchronization details,
it does not contain enough information, such as an history of
modifications and deletions, to properly deal with conflicting
scenarios. Therefore, we cannot fully depend on the native file
systems for tacking synchronization decisions.

Consequently, each {\protbox} instance uses PReg
for this task. For each {\PPair} the PReg stores its Pair
Key and information to detects differences between
the Pair's {\ProtF} and {\SharedF}: (i) encrypted and
decrypted names of each file of the Pair, (ii) last
modification date, (iii) file length and (iv) file's cleartext
contents. For the Pair's directories only the names are stored.

It should be possible to have two or more {\PPair} for the
same {\ProtF}; it enables to share the same {\ProtF} through
several cloud providers, using a different {\PairKey} for each of
them. However, the contrary should be impossible: a single {\SharedF}
cannot be used by more than one {\PPair}. The reason
for this is that all the contents of a {\SharedF} must be
protected with a single {\PairKey}.

Each encrypted file name results from the encryption of the original
name with the {\PairKey}, encoded in a modified Base64 alphabet. In
this alternative alphabet we replaced the ``/'' symbol, which is
very often used as a path separator, by ``-'' (hyphen). For the
encryption/decryption operations we chose the 
ECB mode 
and PKCS \#5 padding. The padding helps
to hide the real length of the original name.

{\protbox} monitors the folders of each Pair to detect
modifications relatively to the PReg information. When the
modification corresponds to a file insertion in one of the folders,
a new entry is inserted in PReg and the file is replicated in the
other folder with the appropriate encryption or decryption
transformation.

When the modification corresponds to a removal of a file or
directory, they are similarly removed from the other folder but not
from PReg, where they become marked as hidden. Furthermore, in the
case of files removed in the {\SharedF}, the cleartext replica in
the {\ProtF} is stored in PReg along with the hidden entry prior
to removing them from the file system, thus enabling {\protbox} to
restore them afterwards upon a user request.

When the modification corresponds to an update of a file in only one
of the Pair's folders, in practice for {\protbox} it corresponds to
a combined removal of the file and insertion of a new one in that
folder. Upon both these steps, all the parties sharing the same
{\CloudF} with {\protbox} that receive a new encrypted version of
a file will store the old (cleartext) replica in their PReg.

When the modification corresponds to an update of the same file in
both folders of a Pair, then we have a conflict. In this case,
{\protbox} renames the file in the {\ProtF} to include the name 
of the local user. Then, it considers both the renamed
file and the updated file in the {\SharedF} (which are no longer
linked) as independent file insertions. The overall, distributed
outcome of this operation may not be always the same, since several
{\protbox} instances may compete in this process, in different
hosts, without central coordination. Nevertheless, no files are lost,
since these files are never deleted by {\protbox}.

\subsection{Identification and authentication of users}

For supporting well-informed decisions by {\protbox} users to
respond positively to {\ProtF} key requests we had to choose a
method for identifying and authenticating {\protbox} users. We
decided to use national eID tokens to achieve both goals, by using
their X.509 authentication certificates and their public keys to
validate signatures on {\PairKey} requests and responses, signed with
the corresponding private keys. 

By using national eIDs, the access to protected files shared through
the cloud only occurs after a two-factor authentication: the
possession of the eID token and the knowledge of a personal
identification number (in order to unlock the token's cryptographic
functionalities). This way, the risk of personification by others,
namely cloud storage providers, is dramatically decreased.

In our protection model we didn't consider the hiding of the users'
identity, expressed in signatures performed with their eID, from the
cloud storage providers. Therefore, these providers can obtain the
real identity of the persons exchanging secure files with
{\protbox}. Dealing with this security issue is a topic for future work.

\subsection{Key Distribution}

For the distribution of {\PairKeys} to individual persons sharing the
same {\CloudF} via {\protbox} we designed a protocol based on the
exchange of special files through the {\CloudF}. These special
files, which are not engaged by {\protbox} synchronization functions,
are identified by starting with ``\_'', which does not belong to
our modified Base64 alphabet.

A {\protbox} instance places a {\PairKey} request in the {\CloudF}
when it needs it to properly decrypt the contents of a related
{\SharedF}. The request contains an encryption public key
(belonging to a \textbf{Key Distribution Key Pair}, \textbf{KDKP})
signed by the requester. This public key should be used by anyone
knowing the {\PairKey} to send it back to the requester. The
signature is made with the eID authentication private key, and the
corresponding certificate should go along with the request.


A {\PairKey} request file has a name that is formed by a leading ``\_''
and an hexadecimal representation of a 128-bit random number. This
number is generated by the requester and will be used to match the
{\PairKey} response. Several persons can place simultaneous requests
in the same {\CloudF}, the probability of collision is nearly
null. A {\PairKey} response will have a similar file name, but with
an additional extension formed by an hexadecimal representation
of another 128-bit random number. This number is generated by the
responder and allows many persons knowing the {\PairKey} to
respond without colliding.

Whenever a {\protbox} instance detects a {\PairKey} request in a {\SharedF}
for which it knows such key, it checks the request signature
and presents the identity of the requester to the local user,
prompting for key distribution authorization. Upon the user
authorization, the {\PairKey}
is ciphered with
the requester public key and the response is signed with the eID
of the responder. The goal of this signature is twofold: (i) it
allows the requester to know who provided the {\PairKey} and (ii) it
prevents anonymous attackers from injecting tampered responses in
the {\CloudF}. Note that we cannot prevent Denial-of-Service
(DoS) attacks against the key distribution protocol (attackers may
be able to tamper or delete {\PairKey} requests and responses), but we
can prevent {\protbox} users from being mislead by anonymous
attackers providing wrong {\PairKeys}. We can still have attacks
providing wrong {\PairKeys}, but since the responses are signed, they
are not anonymous.

{\PairKey} responses are signed tacking into account the request,
i.e., the signature is made upon a hash including the original
request (file name and contents). This way, responses cannot reused,
which is advisable to prevent users to be fooled by replayed responses.

{\PairKey} requests and the corresponding responses are deleted
upon successfully processing a
response. It may happen, however, that some responses may be placed
in the {\CloudF} after the deletion of the request. In this case,
lost responses (easily detected because they have no counterpart
requests) can be deleted by anyone sharing the {\CloudF} after an
acceptable timeout upon detection of the incoherence.

In our protection we didn't consider any mechanism to revoke
accesses to files in a {\SharedF}. Ultimately, this needs to be
explored at the sharing service provided by the storage
provider. Furthermore, we assumed that each person with access to
a {\SharedF} can provide a {\PairKey} response to a key pair
request for that folder. More restrictive response politics (e.g.
only one participant is allowed to respond) must be managed at a
higher level with some form of personal agreement.
Dealing with such policies is a topic for future research.

\subsection{Management of file content restoration}

As previously mentioned, each time a {\SharedF} file
is updated, the corresponding file in the {\ProtF}
is updated accordingly and a
backup copy of the replaced file contents is created. With this
basic behavior, files shared among several users by means of
{\protbox} that go through many small updates are likely to create
long lists of backup contents in many {\protbox} instances.

To deal with this issue, {\protbox} instances offer different
policies for managing the backup of updated files, such as: (i)
never keep a backup copy (ii) limit the number of backups to a
maximum number of copies (iii) ask the user each time a backup copy
is to be made. Because files have different 
relevancy levels,
these policies can be deployed on a per file basis. 


\section{Prototype implementation and experience}
\label{experiences}

A prototype implementing all the features specified in the
architecture was developed using Java, and is available as
a open-source project at \\
\url{https://github.com/edduarte/protbox}
. Aspects like Java's native file 
system recognition were used in order to emphasize maximum
compatibility. Moreover, because it runs on any implementation of
the Java Virtual Machine, it is compatible with popular operating
systems, such as Windows, Mac OS X and Linux. Licensed third-party
libraries that were used for the development of this prototype
(SwingX, ImageJ, JGoodies, Apache Commons, Guava) are all freely
distributed and open-source.

Each {\protbox} instance uses a different PReg for each local user
and uses the user's home directory to store it. The PReg is formed
by a directory for storing backed up files and an encrypted file
containing a serialized Java data structure with all the
user's {\protbox} metadata. This file is encrypted with AES in ECB
mode with a key derived from a user password. This file keeps the
user's KDKP and the random identifiers used in his {\PairKey}
requests; KDKP is generated by {\protbox} on the first execution.

{\PairKey} requests are produced by {\protbox} instances at most
once on each run, since they can be reused for different {\SharedFs}
(while stored in request files with different names, for preventing response
replay attacks). This way, the user signature with his eID for producing a
{\PairKey} request is required at most once each time his {\protbox}
runs. Note, however, that the user may be asked to make other
signatures with his eID, namely for producing {\PairKey} responses.

During start-up, our prototype checks for configuration
files added by the user, which should specify the local path of a
eID token PKCS\#11 provider and the alias of the authentication certificate
contained within said eID token. With this, {\protbox} allows
dynamic support for any national eID token to sign {\PairKey} requests
and responses.

Protected files always start with an integrity control value. After that,
they may contain an optional initialization vector for an encryption
mode (e.g. CBC). Finally, they have the actual file contents
encrypted. The cipher algorithms used for protected files are
defined independently for each {\SharedF}. The person that decides
it is the same that defines its {\PairKey}, which is the first that
creates a protected file on it.

{\PairKey} requests' and responses' signatures contain the complete
certificate chain of the signer's certificate, excluding the root
certificate. This facilitates the validations of the signatures, at
the cost of adding more data to those files. But since they are
transient, this is not an issue. Besides the {\PairKey}, a response
also contains the names of the cryptographic algorithms that are being
used to protect the files in the {\SharedF}.  For such names we
used the strings that are actually used to instantiate cipher
objects using the Java Cryptography Algorithm factory model.
Examples of such strings are ``\texttt{AES/CBC/PKCS5Padding}'' for a
symmetric encryption cipher and ``\texttt{HmacSHA}'' for computing
an integrity control value.

%

The prototype was tested in Windows 7, Ubuntu 12.04.4 and Mac OS X 
Mavericks 10.9.4 operating systems with four of the 
current major cloud storage providers on the market: Dropbox, 
Google Drive, Microsoft OneDrive and SugarSync. Multiple tests were
performed to check if the provider's synchronization methods
displayed considerable loss of performance, since file encryption
is known to interfere with the provider's synchronization
techniques~\cite{tests}. Tests included (i) the creation of a single and of
multiple files in a {\ProtF},
(ii) sharing of a {\PairKey} between several persons, (iii) simultaneous creation
of files in different {\ProtFs} in different hosts for the same {\CloudF}, (iv) detection of tampered files in {\SharedFs}
and (v) file deletion detection and (vi) recovery of deleted files.
Unfortunately, it is impossible to describe here all the interactions
with the users that are triggered within many of these tests.

Under normal conditions, the prototype executed every task
successfully with all of these providers and presented no
distinguishing differences between them in terms of behavior. Under
conditions where the cloud storage service's permissions features
could be set, when reducing the users' permission from ``read/write'' to
``read-only'', {\protbox} could not cope with it, since it could not
even post a {\PairKey} request in the {\SharedF}. However, since
{\protbox} is by design agnostic from specific features provided
by cloud storage providers, such as this file protection mechanism, this is an expected limitation.



\section{Related Work}
In this section we will give an overview of features present in
other existing cloud storage security solutions, 
implemented as third-party software applications, and effectively
compare the overall operations and design of these against {\protbox}.
The analyzed solutions are
BoxCryptor\footnote{\url{https://www.boxcryptor.com/en/boxcryptor}},
Viivo\footnote{\url{http://www.viivo.com}},
CloudFogger\footnote{\url{https://www.cloudfogger.com/en/}},
Sookasa\footnote{\url{https://www.sookasa.com/}},
TrueCrypt\footnote{\url{http://www.truecrypt.org/}} and
CCE (Citizen Card Encrypted)~\cite{cce}.


Similarly to {\protbox}, all of these solutions encrypt files
from the installed cloud service with locally generated 256-bit AES keys.
As an added effort, {\protbox} implements integrity checking of encrypted
files to prevent files with garbage from being produced in {\ProtFs} of
peers. This feature could not be found in the documentation of any of the
analyzed solutions, though it may be in place.


With BoxCryptor and TrueCrypt, plaintext replicas of encrypted files
are maintained in a local virtual disk drive that is created
in the user host, which requires a
strong integration with the operating system kernel of the user
machine. Other solutions, such as Viivo and Sookasa, detect a set of well-known,
locally installed cloud storage providers, and are
limited to encryption of a single folder (and its sub-files and
sub-folders) at the target cloud storage
service (naturaly, the local cloud replica). In contrast, {\protbox}
integrates in a transparent way with the native file system and
prompts the user to freely specify the cloud replica and prot folders
that define a {\PPair}, making it a more intuitive and flexible
solution. In addition, this flexibility allows the configuration of
multiple Pairs based on the same {\ProtF} and different cloud
replicas, introducing simultaneous synchronization 
and encryption of contents into multiple {\CloudFs},
a feature that is not available in other works.

In regards to local protection, CloudFogger and Sookasa do not
replicate files between two different local
folders, instead encrypting and decrypting cloud
folder files on-the-fly according to their actual local usage.
Local files, placed at the cloud folder, are always encrypted,
and are only decrypted to plaintext to the user
when the user authenticates himself within the provided
application
\footnote{\url{http://support.cloudfogger.com/index.php?/\\Knowledgebase/Article/View/10/7/how-secure-is-cloudfogger}}
\footnote{\url{https://sookasa.zendesk.com/hc/en-us/articles/200045197-How-do-I-encrypt-files-}}.
TrueCrypt's keeps a local mountable file with the encrypted
files, which contents can only be accessed when TrueCrypt is
running. 
{\protbox}, like BoxCryptor and Viivo,
keeps the decrypted view (prot folder) available
locally at all times, and because the established objective was
to just protect files residing in the cloud folder,
it does not have any local protection measures on the prot
folder.

For the majority of these solutions, encryption keys and the sharing
logic of these is handled within a backend platform available in a web server.
Users must implicitly trust web server's safeguard.
For example, in BoxCryptor, file sharing is targeted to individual files,
where a random key is generated to encrypt
every different file that can be shared with another single
user or with a group of users. This key is then stored
in the BoxCryptor Key Server and made accessible to the intended
user or group. As security measures, these keys are encrypted with
cryptographic material generated from the user's credentials and
stored locally\footnote{\url{https://www.boxcryptor.com/en/technical-overview#anc02}}.
The remaining solutions allow the sharing of whole directories with
specific users, generating an encryption key per directory and storing
it in the application's backend servers, with access limited to
those users. The encryption material relevant to file
protection is said to be kept locally, without ever being transferred
to these backend servers. This claim cannot be verified because they
are not open-source.
{\protbox} also bases
its file encryption on whole directories, allowing the setting up
of multiple simultaneous Pair sharing, but by structuring a whole
key distribution protocol by transferring files in the shared folder,
{\protbox} avoid the need to implement a sharing and encryption logic
in a backend service.

The authentication paradigm for all of these solutions, except
for CCE, is
knowledge-based, using character-based credentials.
These credentials identify different users
and allow intuitive sharing of files, where a user can specify who
should obtain access to encrypted files by stating the corresponding
accounts. {\protbox}, in contrast, uses a strong ownership-factor
authentication method based on national eID tokens to identify
different users during sharing operations.

The usage of eID tokens for authentication in cloud-security
was already used by CCE~\cite{cce}. CCE implemented
file confidentiality by using the token's provided encryption and
decryption mechanisms, which also means that they are dependent
on these mechanisms being supported by the eID token. Many eID
tokens, like the Portuguese Citizen Card, currently do not support
such capabilities, hence cannot provide file confidentiality
on their own. {\protbox} does not rely on national eIDs having these
features. Instead, it only requires signature capabilities
to allow verification of human identity by peers.

For a more controlled sharing protocol,
Viivo proposes a mediator-based implementation where every shared folder
has a user with
'moderator' privileges, which, by default, is the first
user to attain access to said folder. New users must
request for permission of access to the encrypted contents directly to
the moderator, and this moderator must constantly check for and manage
these requests. Since there is
only one moderator per folder, this moderator must be familiar with
all of the requesting users. With this, a user that is only known as
trustworthy by single or a few users of the shared folder excluding
the moderator will, more likely than not, have his request denied.
With {\protbox}, every single sharing request is sent in a ``multicast"
fashion, and the requirements for one of these request to be accepted is
to provide a valid certificate chain and a valid signature and to have
at least one user accept such a request. The reason for our policy
is that we don't have a central authority for controlling ownership
rights over {\SharedFs}; everyone that has access to the {\SharedF}
is a peer with equal rights.





Finally, all of the available solutions place complete trust in the cloud
storage provider's capacity of protecting files from illegitimate or
unwanted deletions and the capability of backing up files to allow
eventual restore of file contents. 
With that said, and in tune with {\protbox}'s intended agnosticism,
intentional or accidentally deleted files can be recovered
without depending on the cloud service provider's own mechanisms.

\section{Conclusions}

This paper proposed {\protbox} as a multi-platform solution for
cloud storage security, where data confidentiality and sharing is
performed with agnosticism
and autonomy towards the cloud storage service provider. It
adapts to the native file system and to existing cloud
storage services without trusting nor requiring
their capabilities other than the process of synchronization
to the cloud. Features that might already be implemented by providers
like data recovery are also provided by {\protbox} as an independent
and secure way of restoring content without the provider's
acknowledgement.

Regarding other similar existing solutions, {\protbox} does not store
and manage user credentials
in a central or distributed service in order to provide key sharing
functionality. Instead, it uses the cloud environment and the synchronization
of files on {\SharedFs} to enable peer-to-peer exchange
of cryptographic material. In addition, while other solutions
use convenient but weak password-based authentication
measures to identify users, {\protbox} uses a strong eID-based
identification and authentication
paradigm exploring national eID tokens. The end result of this is that,
while other cloud security solutions rely on transferring trust from
the storage service systems to their own systems, with {\protbox} the
user does not need to trust any additional services other than
existing PKI infrastructures for eID exploitation.




Finally, {\protbox} protects the files shared through cloud storage
from being deleted or damaged intentionally or accidentally. This
is achieved by keeping local backup copies of files modified upon
a modification triggered by the cloud provider.

\bibliography{protboxPaper}

\end{document}